\documentstyle[referee,doublespacing,epsf]{mn}

\newcommand{\kms}{\mbox{km s$^{-1}$}}            
\newcommand{\nh}{\mbox{NH$_3$}}                 
\newcommand{\solmass}{\mbox{M$_\odot$}}         

\def\epsfsize#1#2{\ifnum#1>\hsize\hsize\else#1\fi}

\begin{document}


\title[Ammonia observations of MBM 12]
{Ammonia observations of the nearby molecular cloud MBM 12}

\author[J. F. G\'{o}mez et al.]{Jos\'{e} F. G\'{o}mez,$^1$
Joaqu\'{\i}n Trapero,$^{1,2}$
Sergio Pascual,$^3$
Nimesh Patel,$^4$\newauthor
Carmen Morales,$^1$
Jos\'{e} M. Torrelles$^{5,6}$\\
$^1$Laboratorio de Astrof\'{\i}sica Espacial y F\'{\i}sica
Fundamental, INTA, Apdo. Correos 50727, E-28080 Madrid, Spain\\ 
$^2$Universidad SEK, Cardenal Z\'{u}\~{n}iga s/n, Segovia, Spain\\
$^3$Departamento de Astrof\'{\i}sica, Universidad Complutense,
E-28040 Madrid, Spain \\
$^4$Harvard-Smithsonian Center for Astrophysics, 60 Garden St., Cambridge,
MA 02138, USA\\
$^5$Institut d'Estudis Espacials de Catalunya (IEEC/CSIC), Edifici Nexus,
c/ Gran Capit\`{a} 2-4, E-08034 Barcelona, Spain\\
$^6$Instituto de Astrof\'{\i}sica de Andaluc\'{\i}a, CSIC, Apdo. Correos
3004, E-18080 Granada, Spain} 


\maketitle

\begin{abstract}

We present \nh(1,1) and (2,2) observations of MBM 12, the closest
known molecular cloud (65 pc distance), 
aimed to find evidence for on-going star
formation processes. No local temperature (with a $T_{\rm rot}$ upper
limit of 12 K) nor linewidth enhancement is
found, which suggests that the area of the cloud we mapped ($\sim 15'$
size) is not
currently forming stars. Therefore, this close 
``starless'' molecular gas region is
an ideal laboratory to study the physical conditions preceding
new star formation. 

A radio continuum source was found in Very Large Array 
archive data, close but outside the \nh\ emission. This source is
likely to be a background object.
\end{abstract}


\section{Introduction}

MBM 12 (Magnani, Blitz, \& Mundy 1985) is the closest known 
molecular cloud. 
Its distance has been estimated
to be
$\sim 65$ pc (Hobbs, Blitz, \& Magnani 1986). 
This means that the distance to MBM 12 
is less than half of that to the
closest star-formation complexes (e.g., 140 pc to Taurus). 
Therefore, determining the presence of star-forming processes in MBM 12
would allow us to get deep into the smallest possible physical scales
of these processes. For instance, for young low-mass stars a critical scale is
$\sim 100$ AU, which is the estimated size of protoplanetary discs, and the
distance at which jets are being collimated \cite{rod89}. 
A size of 100 AU 
 at 65 pc will extend $\sim 1\farcs 5$, which is a resolution 
attainable at
 present, for instance by millimetre interferometers already
 operating. However, similar studies in more distant star-forming regions
 would require sub-arcsecond resolution, for which we will have to wait
 until the next generation of interferometers is available. 

Although no ongoing star-formation has been detected so far within this
high-latitude cloud, the presence of T Tauri stars \cite{pou96}
outside, but close to
the bulk of
the CO emission 
(Pound, Bania, \& Wilson 1990; Zimmerman \& Ungerechts 1990; 
Moriarty-Schieven, Andersson, \& Wanner 1997), 
suggests that, at least in the near past,
this cloud has been a star-forming site. However, it might be possible that
these T Tauri stars are run-away objects from other star-forming
regions. The detection of embedded young stellar objects would then be the
only way to confirm whether MBM 12 is actually a star-forming region. 

Reach et al.\ \shortcite{rea95} have determined the presence of dense gas 
($\ga 10^4$ cm$^{-3}$, 
traced by CS emission) 
in some areas of MBM 12, forming clumps in near virial
equilibrium (i.e., dynamically bound, 
not just density fluctuations in a turbulent medium). 
Therefore, these are possible pre-stellar cores (Ward-Thompson et
al.\ 1994) that merit
further study. 

In this paper, we present single-dish \nh\ observations aimed to
search for evidence of on-going star formation in MBM 12. 
\nh\
observations can be used to identify the presence of current star
formation processes.
The ratio between different \nh\
transitions give us temperature information (Ho \& Townes 1983). 
Local heating effects
traced by ammonia have been successfully used to locate young stellar
objects (e.g., G\'{o}mez et al.\ 1994; Girart et al.\ 1997). 
Moreover, local peaks of linewidth could indicate
local turbulence enhancements, which are also very likely associated
with star-forming processes.

We also used radio-continuum observations 
from the Very Large Array (VLA) of 
NRAO\footnote{The National Radio
Astronomy Observatory is a facility of the National Science Foundation
operated under cooperative agreement by Associated Universities, Inc.}
archive,
to look for possible embedded
young stellar objects.

\section{Observations}


Observations of the \nh(1,1) and (2,2) inversion lines
(rest frequencies $\nu_{11} = 23.6944955$ GHz, and $\nu_{22} =
23.7226333$ GHz) 
were taken with the 37m antenna at 
Haystack Observatory\footnote{Radio Astronomy at MIT
Haystack Observatory is supported by the US National Science
Foundation.} on
1998 March 17-24, and 1998 May 13-15. The beam size of the telescope
at this frequency is $\sim  1\farcm 44$ (which corresponds to $\sim
0.03$ pc on MBM 12). We used a dual maser,
cryogenically cooled receiver, which was tuned at a frequency midway
between those of the observed lines, and with $V_{\rm LSR}= -5.0 $ \kms. 
The autocorrelation spectrometer
was configured to measure a bandwidth of 53.3 MHz with 1024 lags, which
gave a spectral resolution of 62.85 kHz (0.8 \kms). 
This system setup allowed us to obtain spectra that
included both \nh\ lines. Two circular polarizations were also
obtained simultaneously. The rms pointing error was estimated to be
$\sim 11''$, with observations of Saturn.
The reference position of our \nh\ maps is 
$\alpha(1950) = 02^h 53^m 08\fs9$, $\delta(1950) = +19^\circ 14' 15''$.
The spectra have all been corrected for antenna gain and atmospheric
attenuation. Typical system temperatures are $\sim 100$ K. 
The final spectra have been smoothed up to a velocity
resolution of 1 \kms, yielding an rms noise level of $\sim 0.03$
K.
Data reduction was performed with the program CLASS of IRAM and
Observatoire de Grenoble.

We also used archive VLA data  towards this region. These data
correspond to B-array, 3.6 cm continuum observations, originally taken
by D. Helfand and J. Halpern on 1990 Aug 2, to search for a radio
counterpart of the X-ray pulsar H0253+193 (see Patterson \& Halpern
1990; Zuckerman et al.\ 1992). 
An effective bandwidth of 100 MHz, and
two circular polarizations were observed. The phase centre was located
at $\alpha(1950) = 02^h 53^m 20\fs5$, $\delta(1950) = +19^\circ 14'
38''$. The source 1328+307 was used as the primary flux calibrator,
with an assumed flux density of 5.19 Jy. The phase calibrator was
0235+164, for which a flux density of $3.44\pm 0.06$ was
derived. 
Calibration and further image processing was performed with
the program AIPS of NRAO. The resulting synthesized beam was 
$0\farcs 84\times 0\farcs 72$,  and the rms noise level $1.0\times
10^{-2}$ mJy beam$^{-1}$.

\section{Results} 
\begin{figure}
\epsffile{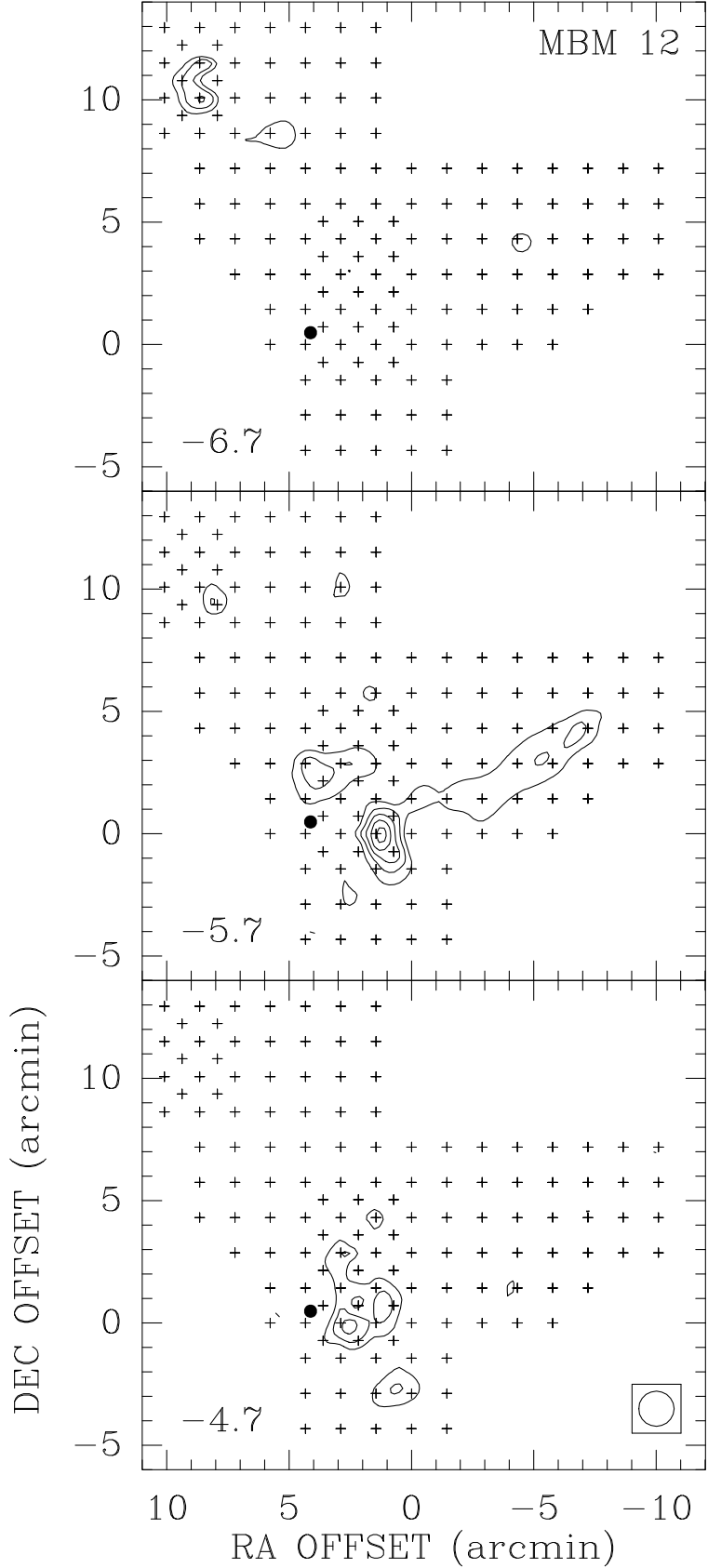}
\caption{Contour maps of NH$_3$(1,1) intensity, at different
velocities. The LSR velocity (in \kms) 
is indicated at the bottom left corner of
each panel. The lowest contour is 0.09 K and the increment step is
0.03 K ($1\sigma$). The filled circle indicates the position of the
radio source VLA B0253+192. Crosses indicate the observed points. The
beam size is indicated as a circle at the bottom right corner.}
\label{vmaps}
\end{figure}

\begin{figure}
\epsffile{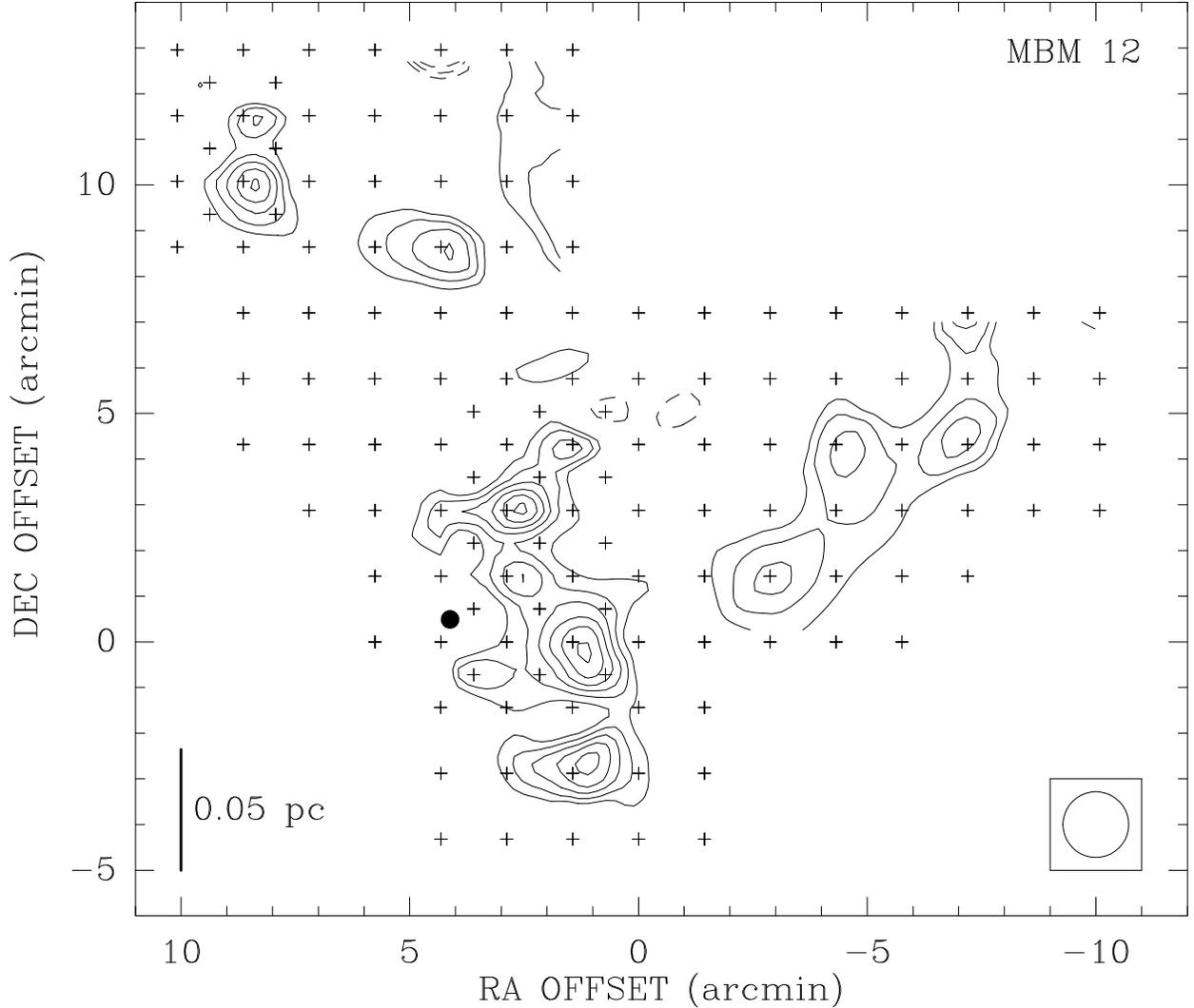}
\caption{Contour map of the integrated emission of the
\nh(1,1) main hyperfine component. The integration range was between
$-8.1$ and $-2.9$ \kms. The lowest contour map is 0.15 K \kms, and the
increment step 0.05 K \kms. 
Symbols have the same meaning as in Fig.~\ref{vmaps}.}
\label{integ}
\end{figure}

\begin{figure}
\epsffile{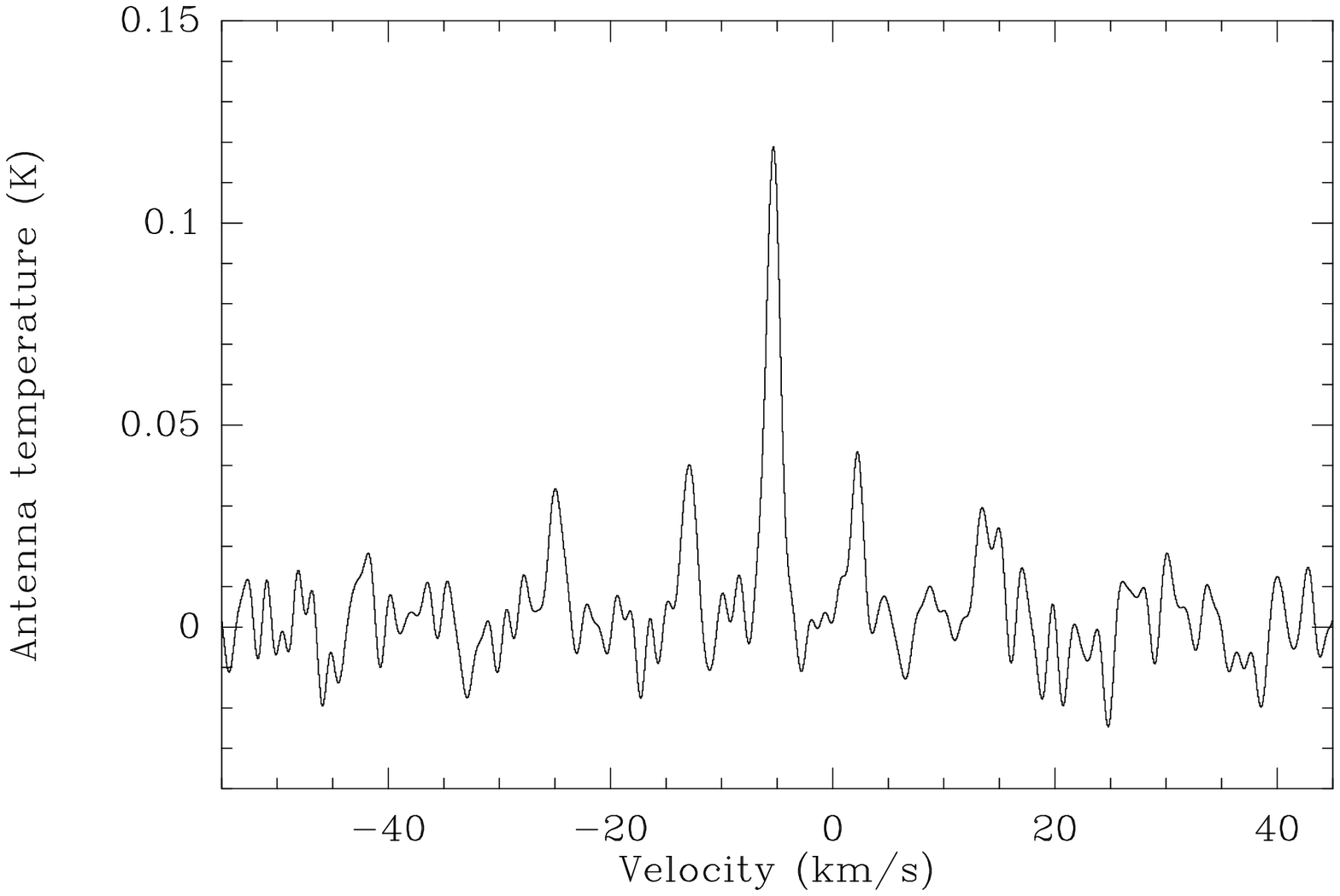}
\caption{Averaged spectrum of the \nh(1,1) line, within a box of
$3'\times 3'$ around position offset $(1\farcm44,0\farcm00)$.}
\label{spec}
\end{figure}

Figure \ref{vmaps} shows the emission of the \nh(1,1) line detected at
different velocities. Figure \ref{integ} shows the integrated intensity
map. Several clumps of ammonia emission are evident in this latter figure, 
the most intense one centred around position offset
$(1\farcm44,0\farcm00)$. The observed clumps tend to align forming two
clear filamentary structures, one extending south to north between
position offsets $(1\farcm44,-2\farcm88)$ and $(1\farcm44,4\farcm32)$,
and the other one from southeast to northwest between
$(-2\farcm88,1\farcm44)$ and $(7\farcm20,4\farcm32)$. The morphology
seen with our \nh\ maps is similar to the one in CO
(Pound et al.\ 1990; Zimmerman \& Ungerechts 1990; 
Moriarty-Schieven et al. 1997), 
and specially to the CS map \cite{rea95}.
Figure \ref{spec} shows an averaged spectrum
within a box of $3' \times 3'$ around $(1\farcm44,0\farcm00)$, which
includes the \nh\ emission from the most intense clump. Apart from the
central, main hyperfine component, the satellite lines of \nh(1,1) are
also evident.
No \nh(2,2) line emission was detected in our spectra. This emission
is not present even in the
average within the $3' \times 3'$ box around $(1\farcm44,0\farcm00)$, with an upper
limit\footnote{Upper limits and errors
are given in this paper for a 99\% 
confidence level.} of $0.020$ K.

No radio continuum emission is detected within the bounds of the
ammonia emission, with an upper limit to the flux density of
$2.3\times 10^{-2}$ mJy. Only an unresolved source is present within the
primary beam of the VLA, with a flux density of $0.114\pm0.026$
mJy, at position $\alpha(1950) = 02^h 53^m 26\fs38$, $\delta(1950) =
+19^\circ 14' 44\farcs 4$ (we name this object as VLA B0253+192). 
Note that the phase centre of the VLA
observations is located at position offset (2\farcm74,0\farcm38) in our \nh\ maps,
and the primary beam at 3.6 cm is $5\farcm4$ (FWHM). Therefore, this primary
beam includes the whole central clump. The position of VLA B0253+192
is indicated in Figures \ref{vmaps} and \ref{integ}.

\section{Discussion and Conclusions}

The main goal of our \nh\ observations was to identify the presence of
on-going star formation processes, 
by measuring local temperature and turbulence
enhancements. 

In our data, the lack of \nh(2,2) emission indicates a low temperature
throughout the mapped area. In particular, from the averaged data
for the most intense clump, around $(1\farcm44,0\farcm00)$ 
(see Fig.\ \ref{spec}), and using the values for the main and inner
satellite components $T_A(1,1;m) =
0.117$ K $\pm 0.022$ K,
$T_A(1,1;is)= 0.041$ K $\pm 0.016$ K, $T_A( 2,2;m) < 0.020$ K, we derive an optical depth of
$\tau(1,1;m) = 0.69$ and an upper limit to the
rotational temperature of $\sim 12$ K. This temperature is similar to
those found in other high-latitude clouds, including MBM 7
\cite{tur95}, which is close to MBM 12. It is also expected for this
rotational temperature to be close to the value of the local kinetical
temperature \cite{tur95}.
No local linewidth enhancement is seen in our data. Typical linewidths
seem to be unresolved with our 1 \kms\ velocity resolution.

From these results we conclude that no indication of current star
formation is evident in our \nh\ data. This suggests that T Tauri
stars seen close to MBM~12 \cite{pou90} 
were formed in a near past, but
star-forming processes are no longer going on, at least in the area we
mapped. Alternatively, those T Tauri stars may be runaway objects from
nearby star-formation sites.

However, Reach et al. \shortcite{rea95} estimated that MBM 12 is a 
potential star-forming cloud,
given the similar mass derived from the CS integrated intensity
and the virial mass. In our data, for the main
clump around $(1\farcm44,0\farcm00)$, and assuming a $T_k=T_{\rm rot}
= 12$ K, we estimate 
a mass of $0.12 \left(\frac{X_{\rm NH_3}}{10^{-8}}\right)^{-1} $
\solmass, where $X_{\rm NH_3} $ is the molecular abundance of \nh\
with respect to H$_2$ . 
The total mass for the clumps mapped 
in our data is
$\sim 0.65 \left(\frac{X_{\rm NH_3}}{10^{-8}}\right)^{-1} $ \solmass. 
These values are consistent with those obtained
by Reach et al. \shortcite{rea95}. 
Although we see no sign of current star formation
processes, the
presence of self-gravitating clumps suggests the possibility of these
processes to occur in the future. Therefore, given its proximity to the
Sun, this region is an ideal
laboratory to study precollapsing physical conditions, preceding new
star formation (e.g., Ward-Thompson et al.\ 1994).

On the other hand, the radio-continuum  source that lie within our
sampled area does not seem to be a young stellar object associated to
MBM 12, since it is outside the bounds of the \nh\
emission. Association with \nh\ emission is a known characteristic of
young stellar objects \cite{ang89}. 
The source VLA B0253+192 is likely to be
an extragalactic object.

\section*{acknowledgments}

We thank John Ball, Kevin Dudevoir, and Phil Shute
for their help during
the observations at Haystack Observatory. JFG and JMT are supported in
part by DGICYT grant PB95-0066 and by Junta de Andaluc\'{\i}a (Spain).
SP is supported by DGICYT grant PB96-0610.



\begin{thebibliography}{}

\bibitem[\protect\citename{Anglada et al.\ }1989]{ang89}
Anglada, G., Rodr\'{\i}guez, L.F., Torrelles, J.M., Estalella, R.,
Ho, P.T.P, Cant\'{o}, J., L\'{o}pez, R., Verdes-Montenegro, L., 1989,
ApJ, 341, 208

\bibitem[\protect\citename{Girart et al.\ }1997]{gir97}
Girart, J.M., Estalella, R., Anglada, G., Torrelles, J.M., Ho, P.T.P.,
Rodr\'{\i}guez, L.F., 1997, ApJ, 489, 734

\bibitem[\protect\citename{G\'{o}mez et al.\ }1994]{gom94}
G\'{o}mez, J.F., Curiel, S., Torrelles, J.M., Rodr\'{\i}guez, L.F.,
Anglada, G., Girart, J.M., 1994, ApJ, 436, 749

\bibitem[\protect\citename{Ho \& Townes.\ }1983]{ho83}
Ho, P.T.P., Townes, C.H., 1983, ARA\&A, 21, 239

\bibitem[\protect\citename{Hobbs et al.\ }1986]{hob86}
Hobbs, L.M., Blitz, L., Magnani, L., 1986, ApJ, 306, L109

\bibitem[\protect\citename{Magnani et al.\ }1985]{mag85}
Magnani, L., Blitz, L.,  Mundy, L., 1985, ApJ, 295, 402

\bibitem[\protect\citename{Moriarty-Schieven et al.\ }1997]{mor97}
Moriarty-Schieven, G.H., Andersson, B-G., Wanner, P.G., 1997, ApJ,
475, 642

\bibitem[\protect\citename{Patterson \& Halpern }1990]{pat90}
Patterson, J., Halpern, J.P., 1990, ApJ, 361, 173

\bibitem[\protect\citename{Pound }1996]{pou96}
Pound, M.W., 1996, ApJ, 457, L35

\bibitem[\protect\citename{Pound et al.\ }1990]{pou90}
Pound, M.W., Bania, T.M., Wilson, R.W., 1990, ApJ, 351, 165

\bibitem[\protect\citename{Reach et al.\ }1995]{rea95}
Reach, W.T., Pound, M.W., Wilner, D.J., Lee, Y.U., 1995, ApJ, 441, 244

\bibitem[\protect\citename{Rodr\'{\i}guez }1989]{rod89}
Rodr\'{\i}guez 1989, in Structure and Dynamics of the Interstellar Medium,
IAU Coll. 120 (Berlin:Springer), 197  

\bibitem[\protect\citename{Turner }1995]{tur95}
Turner, B.E., 1995, ApJ, 444, 708

\bibitem[\protect\citename{Ward-Thompson et al.\ }1994]{war94}
Ward-Thompson, D., Scott, P.F., Hills, R.E., Andr\'{e}, P., 1994,
MNRAS, 268, 276

\bibitem[\protect\citename{Zimmermann \& Ungerechts }1990]{zim90}
Zimmermann, T., Ungerechts, H., 1990, A\&A, 238, 337

\bibitem[\protect\citename{Zuckerman et al.\ }1992]{zuc92}
Zuckerman, B., Becklin, E.E., McLean, I.S., Patterson, J., 
1992, ApJ, 400, 665

\end{thebibliography}
\end{document}